\documentclass{article}[12pt]

\begin{document}

\title{Cosmological model with dynamical curvature}
\author{Peter C. Stichel\\
Fakult\"at f\"ur Physik, Universit\"at Bielefeld\\
D-33501 Bielefeld, Germany\\ 
e-mail: peter@physik.uni-bielefeld.de}
\date{06.04.2016}

\maketitle

\begin{abstract}
The recently introduced relativistic Lagrangian darkon fluid model (EPJ C (2015) 75:9) is generalized to a self-gravitating, irrotational, pressure-less and stress free geodesic fluid, whose energy-momentum tensor is dust-like with nontrivial energy flow, defining the general framework for our cosmological model. In the present paper we consider only the averaged dynamics of the Universe at very large scales,  allowing us to work with  conformal flat geometry. The corresponding covariant propagation and constraint equations are treated in a shear-free nonrelativistic limit. The dynamics is described with a Lagrangian in minisuperspace. Analytic solutions for the energy density, the Hubble function and the 1st order relativistic correction to the spatial curvature are obtained by means of two conditional dynamical symmetries. This leads to a cosmological model where the accelerated expansion of the Universe is driven by a time-dependent spatial curvature without the need for introducing any kind of dark energy. A differential equation for the area distance is derived in addition.
\end{abstract}

\section{Introduction}
There is no doubt about that the present universe undergoes a phase of accelerated expansion (cp. [1]) and the literature cited therein). Almost all observations are in good agreement with the $\Lambda$-cold dark matter (CDM) cosmology [2], however some observations are in disagreement with this $\Lambda$CDM model (see [3]). From the theoretical point of view, no convincing solution for the ``Cosmological constant problem'' is known (cp. [2], [4]). 

Two alternative strategies to the $\Lambda$CDM model are under discussion in the physics community.

In the first category one introduces some kind of ``new physics'' by changing Einstein's field equations (EFEs). The $\Lambda$CDM model belongs to this category, since introducing a cosmological constant modifies the original EFEs. 

In the second category one considers, in the framework of  ``old physics'',  accelerated expansion as an apparent effect due to averaging over inhomogeneities in the Universe (for a recent review see [5]). 

To the first category belong alternative cosmological models which explain the accelerated expansion and/or dark matter by changing either the geometrical part of the EFEs (called modified gravity; see [6], [7] and the literature cited therein), or by changing the matter part by adding some scalar fields (for a recent review see [8]). But all these proposals are of a phenomenological nature, because they contain either unknown parameters or even freely chosen functions. To overcome such an exaggerated freedom, some new (i.e. nonstandard) physics is needed, which, however, should be based on known fundamental physical principles (e.g. symmetry). Such an attempt has been made recently by the present author and by W. J. Zakrzewski [9]. Based on General Relativity (GR) we introduced a Lagrangian model containing no unknown parameters, which led to an energy-momentum tensor consisting of a dust term and a nontrivial energy flow [9].

Thus our model belongs to the class with a modified matter part of the EFEs. The model is a GR- generalization of the nonrelativistic darkon fluid model introduced in [10], further developed in [11, 12] and reviewed and extended in [13]. The guiding principle of the latter model is to use the unextended Galilei algebra, dynamically realized by a fluid consisting of massless Galilean particles minimally coupled to gravity. We called it darkon fluid since it had been introduced initially for the description of the dark sector of the Universe.  As argued in [9], the solution manifold contains also the baryonic matter as dust, though. 

The nonrelativistic model describes successfully the transition from  decelerating to accelerating phase of the Universe, as well as the Hubble function and the flat behavior of galactic rotation curves [10, 13]. However, the nonrelativistic framework did not allow us to discover the physical mechanism which is behind the transition from a decelerating to an accelerating phase. 

Initially, we tried to understand our results in terms of some dynamical dark energy (repulsive gravity determined by a negative pressure) [13].  However this is misleading, as we have learned by now from the relativistic treatment.  We found that at sub-Hubble scales the relativistic model reproduces the nonrelativistic results, but with a new physical interpretation:  The present-day  cosmic acceleration is not attributed to any kind of dark energy (negative pressure), it is due rather to a dynamically determined negative spatial curvature [9].  But due to the restricted validity of the sub-Hubble scale approximation, we could not use this interpretation up to or beyond the transition point between decelerating and accelerating phases. In the present paper we extend this result as far as a nonrelativistic description of the Universe is permitted. 

In section 2 we start  with a Lagrangian-free generalization of our relativistic model [9] which contains, as ingredient, a self-gravitating, irrotational, pressure-less and stress free geodesic fluid. The dynamics is determined by an energy-momentum tensor consisting only of an energy density term and of a nontrivial energy flow vector. This defines the general framework for our cosmological model. 

In the present paper we consider only the averaged dynamics of the Universe at very large scales which, allows us to work with conformally flat geometry.  In section 3 we discuss the corresponding covariant propagation and constraint equations for the kinematic quantities. In section 4 we tackle first the problem of vanishing or small fluid shear. This limit turns out to be discontinuous. The two limits, namely the shear-free and the nonrelativistic limit, can not be interchanged. So the shear-free nonrelativistic limit has to be defined by taking the two limits in a definite way simultaneously. We will show that the leading terms of our equations in this limit contain a nontrivial and computable 1st-order relativistic contribution to the spatial curvature (which, however, vanishes to leading order). In section 5 we state the cosmological equations. In section 5.1 we derive them  as Euler-Lagrange Equations in minisuperspace and then we consider their conditional dynamical symmetries as well as their dilation symmetry,  derived as a residual symmetry after gauge fixing. In section 5.2 we derive from these dynamical symmetries the analytic solutions of the cosmological Equations. These are already known from our former work [10]-[13], but now they are given in terms of the dynamically determined spatial curvature.

Due to the non-vanishing energy flow we cannot use the FLRW-formula for the angular diameter distance $D_A$. In section 6 we derive the differential equation to be obeyed by $D_A$ for our model, as well as analytic expressions for the corresponding coefficients.  Some concluding remarks are presented in section 7.

\section{Assumptions}
Within the framework of GR we consider a self-gravitating fluid, described by a velocity field $u^\mu (x)$
($u^\mu u_\mu = - c^2$; $c$ denotes the velocity of light; Greek indices run from 0 to 3 and we use the usual summation convention)
with the following properties:
\begin{description}
\item{$\bullet$} The fluid flow is geodesic (vanishing acceleration)
\begin{equation}
\dot{u}^\mu : = u^\nu \nabla_\nu u^\mu = 0,
\end{equation}                                                                                           
where $\nabla_\nu$ is the covariant derivative given in terms of a torsion-free connection (Christoffel symbols).
\item{$\bullet$} The fluid flow is irrotational. So, with (1), the covariant derivative of $u_\mu$ is decomposed as,
\begin{equation}
\nabla_\nu u_\mu = \sigma_{\mu\nu} + \frac{1}{3} h_{\mu\nu} \Theta,
\end{equation}
where $\sigma_{\mu\nu}$ is the symmetric and trace-less shear tensor, $h_{\mu\nu} : = g_{\mu\nu} + c^{-2} u_\mu u_\nu$                                                                             
is the spatial  projector w.r.t. $u^\mu$ and $\Theta : = \nabla_\mu u^\mu$ is the rate of volume expansion.
\item{$\bullet$} The energy-momentum tensor (EMT) $T_{\mu\nu}$, which represents the right hand side of the EFEs ($\kappa : = 8 \pi G$; $G_{\mu\nu}$                                 denotes the Einstein tensor)
\begin{equation}
G_{\mu\nu} = c^{-4} \kappa T_{\mu\nu}
\end{equation}
is supposed to be pressure-less and stress-free. Therefore the EMT  is decomposed as,
\begin{equation}
T_{\mu\nu} = \rho u_\mu u_\nu + q_\mu u_\nu + q_\nu u_\mu
\end{equation}
where $\rho$ is the energy density in the comoving frame and $q_\mu$ is the energy flow 
vector ($u^\mu q_\mu = 0$).  

The energy density $\rho$ is the sum of the dark sector contribution and of the baryonic contribution.

The physical relevance of the energy flow vector will become apparent later (it determines the time-dependence of the spatial curvature).
\end{description}

All of the above listed assumptions are fulfilled by the relativistic darkon fluid model [9] restricted to spherical symmetric geometry. Hence the fluid model defined by these assumptions is a Lagrangian-free generalization of our relativistic darkon fluid model [9].

\section{Covariant propagation and constraint equations}

Instead of describing the dynamics of our fluid by the EFEs (3) it is more convenient for our purpose to use the equivalent set of propagation and constraint equations for the kinematical quantities $\rho, \Theta, q_\mu$  and $\sigma_{\mu\nu}$ as given e.g. by Ellis and van Elst in their Cargese Lectures 1998 [14] (see also [15]).

In the present paper we will consider only the averaged dynamics of the Universe at very large scales. Now we remind that the FLRW model is conformal flat. Therefore we assume,  for very large scales,  conformally flat geometry (vanishing of the Weyl tensor or, equivalently, vanishing of the electric and magnetic part of the Weyl tensor) also in our case. 

Then, by using the assumptions stated in section 2, we obtain from section 2.2 of [14] (but including the correct factors of c):

\bigskip
\noindent
{\bf 3.1} From the Ricci identities for $u^\mu$ follow

\begin{description}
\item{$\bullet$} the Raychaudhiri-Ehlers equation ($2\sigma^2 : = \sigma_\mu^{~\nu} \sigma_\nu^{~\mu}$)
\begin{equation}
\dot{\Theta} + 2 \sigma^2 + \frac{1}{3} \Theta^2 = - \frac{\kappa}{2} \rho
\end{equation}

\item{$\bullet$} the shear-propagation equation
\begin{equation}
\dot{\sigma}_{\mu\nu} - \frac{2}{3} h_{\mu\nu} \sigma^2 + \sigma_\mu^{~\beta} \sigma_{\beta\nu} + \frac{2}{3} \Theta \sigma_{\mu\nu} = 0
\end{equation}
which, equivalently, may  be written in terms of the 3-space Ricci tensor $^3R_{\mu\nu}$
\begin{equation}
\dot{\sigma}_{\mu\nu} + \frac{1}{3} h_{\mu\nu} (\frac{2}{3} \Theta^2 - 2\sigma^2 - 2\kappa\rho ) + \Theta \sigma_{\mu\nu} = - ~{^3R}_{\mu\nu} c^2 \ . 
\end{equation}                                                                                                               
Contraction of (7) leads to the energy constraint or generalized Friedmann equation
\begin{equation}
{^3R}~c^2 + \frac{2}{3} \Theta^2 - 2 \sigma^2 = 2 \kappa \rho \ .
\end{equation}

\item{$\bullet$} the momentum constraint equation ($D_\mu$ is the covariant derivative in the 3-space with metric $h_{\mu\nu}$)
\begin{equation}
D_\mu \sigma^{\mu\nu} - \frac{2}{3} D^\nu \Theta = -c^{-2} \kappa q^\nu \ .
\end{equation}
\end{description}

Here and in the following an over-dot denotes the covariant time derivative along the fluid flow:  For any tensor field $A$ we define
$$
\dot{A} : = u^\mu \nabla_\mu A
$$

\bigskip
\noindent
{\bf 3.2}	The once-contracted Bianchi identities imply the constraint equations 

\begin{description}
\item{$\bullet$} 
\begin{equation}
D^\nu \rho + \frac{3}{2} c^{-2} \sigma^\nu_{~\mu} q^\mu = c^{-2} \Theta q^\nu
\end{equation}

\item{$\bullet$} 
\begin{equation}
\rho \sigma_{\mu\nu} + D_{(\mu} q_{\nu )} - \frac{h_{\mu\nu}}{3} D_\lambda q^\lambda = 0
\end{equation}
and

\item{$\bullet$} 
\begin{equation}
\epsilon^{\alpha\beta\gamma (\delta} \sigma^{\eta)}_{~\gamma} u_\alpha q_\beta = 0,
\end{equation}
where round brackets indicate symmetrization. 
Equations (11) and (12) are the remainders of the propagation equations for the       
electric and magnetic part of the Weyl tensor.  They guarantee that an initially vanishing Weyl 
tensor will remain zero.
\end{description}

\bigskip
\noindent
{\bf 3.3} From the twice-contracted Bianchi identities we infer

\begin{description}
\item{$\bullet$} the covariant conservation of the EMT                                                                                                                                                       
\begin{equation}
\nabla_\mu T^{\mu\nu} = 0 \ .
\end{equation}
The time-like part of (13) gives the local energy conservation equation
\begin{equation}
\dot{\rho} + \Theta \rho + D_\mu q^\mu = 0
\end{equation}
and the space-like part gives the propagation equation for the energy flow vector
\begin{equation}
\dot{q}_\mu + \sigma_\mu^{~\nu} q_\nu + \frac{4}{3} \Theta q_\mu = 0 \ .
\end{equation}
The solution of (14) for the energy density consists of two parts: A particular solution of the inhomogeneous equation (dark sector contribution) plus a solution of the homogeneous equation (dust-like baryonic contribution).
\end{description}

\section{Nonrelativistic shear-free limit}                       

It is well known that,  for a general perfect fluid, the nonrelativistic limit (NRL) and the shear-free limit are not interchangeable. For the relativistic shear-free perfect fluid either the expansion or the rotation vanishes, whereas no such restriction  exists for the corresponding nonrelativistic fluid (cp. [16] and the literature cited therein).  For our case we obtain an analogous result:

\bigskip
\noindent
{\bf Theorem}

\smallskip
\noindent
For the relativistic shear-free, irrotational, pressure-less and stress-free geodesic fluid the energy flow vector vanishes (cp. [9] for the particular case of spherical symmetry).

For the proof we use of the following results obtained in this context by Coley and McManus [17] (we take $c = 1$):

\begin{description}
\item{$\bullet$} For the shear-free, irrotational and stress-free geodesic  fluid we may use co-moving synchronous coordinates so that the metric has the form (Roman indices run from 1 to 3) [17]
\begin{equation}
d s^2 = -d t^2 + \Omega^{-2} (dx^i)^2
\end{equation}
with
\begin{equation}
\Omega (t,x^i) = a(t) (x^i)^2 + b_i (t) x^i + c(t)
\end{equation}
and $\Theta$ is related to $\Omega$ by
\begin{equation}
\Theta = - 3 \frac{\dot{\Omega}}{\Omega} \ .
\end{equation}
                                                                                                                                                         
\item{$\bullet$} Vanishing pressure leads to the following equation to be satisfied by $\Omega$ [17]    
\begin{equation}
0 = 2 \left( \frac{\dot{\Theta}}{\Theta}\right)^\cdot - 3 \left( \frac{\dot{\Theta}}{\Theta}\right)^2 - \frac{2}{3} \Omega \Delta \Omega + (\partial_i \Omega)^2 \ .
\end{equation}
\end{description}

A lengthy but straightforward computation shows that the general solution of (19) with (17) is given by  
\begin{equation}
\Omega = c(t) (kr^2 + \beta_i x^i + 1) \ ,
\end{equation}
where $k$ and $\beta_i$ are arbitrary constants.
Hence by (18) $\partial_i \Theta = 0$ and therefore, in arbitrary coordinates, (9) leads, for vanishing shear to 
\begin{equation}
q_\nu = 0 \ 
\end{equation}

On the other hand by taking first the nonrelativistic limit and putting afterwards the shear equal to zero will not bother the energy flow vector.                                         
We conclude: The nonrelativistic shear-free limit of the model defined in section 2 is discontinuous.

So let us consider the case when, initially, the shear does not vanish but is almost shear-free  (quasi-shear-free condition) [18]:
\begin{equation}
\sigma (0,\vec{x}) = \epsilon_1 \bar{\sigma} (\vec{x}), \qquad|\epsilon_1| \ll 1 \ .
\end{equation}                                                                                                                                                                  
For the particular case of spherical symmetry  Herrera et al. [18] have shown, by integrating (6), that the shear will remain quasi-shear-free if $\Theta > 0$ (which is always fulfilled in cosmology for an expanding Universe). We guess the stability of (22) will hold in the general case too. Supposed this will not be the case we could restrict the following considerations to the spherically symmetric case. We would not lose anything as the nonrelativistic shear-free limit has this symmetry. 

The NRL is defined by taking the limit $\epsilon_2 \to 0 ~~(\epsilon_2 : = c^{-2})$ in all of the propagation and constraint equations listed in section 3. To be more precise we will take the covariant NRL given by the Newton-Cartan geometry [19] defined by a degenerate metric consisting of a ``time-direction'' $n_\mu$ and a spatial metric $\tilde{h}^{\mu\nu}$ with nullvector $n_\mu$. The corresponding covariant derivative $\tilde{\nabla}_\mu$ is defined by (see (24) for $\tilde{h}^{\alpha\beta}$)
\begin{equation}
\tilde{\nabla}_\mu \tilde{h}^{\alpha\beta} = 0 = \tilde{\nabla}_\mu n_\gamma \ .
\end{equation}
Then the metric and the kinematic quantities show the following limiting behavior [20, 21] 
($\tilde{A}$ denotes the nonrelativistic limit of an arbitrary field or operator $A$):
\begin{eqnarray}
g^{\mu\nu} & \to & \tilde{h}^{\mu\nu} + 0 (\epsilon_2),\nonumber\\
g_{\mu\nu} &\to & - \frac{1}{\epsilon_2} n_\mu n_\nu + \gamma_{\mu\nu} + 0 (\epsilon_2), \quad \nabla_\mu \to \tilde{\nabla}_\nu \ , \nonumber\\
u^\mu &\to & \tilde{u}^\mu, \quad u_\mu \to - \frac{1}{\epsilon_2} n_\mu \ , \nonumber\\
q^\mu &\to & \tilde{q}^\mu ~(\tilde{q}^\mu n_\mu = 0), \quad \rho \to \tilde{\rho}
\end{eqnarray}
hence
$$
\Theta \to \tilde{\Theta} \quad \mbox{and} \quad \sigma_\mu^{~\nu} \to \tilde{\sigma}_\mu^{~\nu} \ .
$$

We remark that the connection $\Gamma$, belonging to the covariant derivative defined by (23), is not unique, let alone in the torsion-free case  [21].  
But for the following it is irrelevant to which element out of the class of admissible $\Gamma$ the relativistic covariant derivative will converge.

To get a meaningful definition for the ``nonrelativistic shear- free limit'' we have to consider an appropriate sequence of relativistic models with non-vanishing shear which converges to a model defined by taking first $c^{-2} \to 0$ ($\epsilon_2 \to 0$) and afterwards $\sigma_{\mu\nu} \to 0$ ($\epsilon_1 \to 0$) in all of the equations in section 3. To achieve this goal we introduce a two-dimensional vector $\vec{\epsilon} = (\epsilon_1,\epsilon_2$) or, in polar coordinates  $\epsilon_1 = \epsilon \cos \alpha, ~ \epsilon_2 = \epsilon \sin \alpha$.                              
Then we define the wanted sequence by taking the limit $\epsilon \to 0$ at fixed $\alpha$ with $0 < \alpha < \pi / 2$.                         
This way we have excluded from our sequence the singular case of taking first  $\epsilon_1 = 0$ and afterwards $\epsilon_2 \to 0$.

From the requirement that the r.h.s. of (7) has a finite NRL we obtain the expansion
\begin{equation}
^3R_{\mu\nu} = \epsilon_2~^3_1R_{\mu\nu} + 0 (\epsilon^2_2)
\end{equation}
yielding  two consequences~:
\begin{description}
\item{$\bullet$} the leading order dynamical equations contain a non-vanishing first order relativistic contribution ${^3_1R}_{\mu\nu}$ to the 3-space Ricci tensor,

\item{$\bullet$} the vanishing of $^3R_{\mu\nu}$ to leading order implies that the 3-space is flat (``in 3-space the Riemann tensor is completely specified by the Ricci tensor'', [15] section 2.7.6). Therefore we may use locally Galilean coordinates defined by [21]
\begin{equation}
n_\mu = \delta_\mu^\circ , ~~~~\tilde{h}^{ij} = \delta^{ij}
\end{equation}
hence
$$
\tilde{u}^\mu = (1,\tilde{u}^i) \ .
$$
\end{description}

Then our propagation and constraint equations given in section 3 read, in the nonrelativistic shear-free limit, in Galilean coordinates (here and in the following we omit the wiggly line at the top of  nonrelativistic fields and $D_t : = \partial_t + u_k \partial_k$ denotes the convective derivative):
\begin{description}
\item{$\bullet$} Raychaudhiri-Ehlers (RE) equation
\begin{equation}
D_t \Theta + \frac{1}{3} \Theta^2 = - \frac{\kappa \rho}{2}
\end{equation}

\item{$\bullet$} Generalized Friedmann equation 
\begin{equation}
^\ast\!R + \frac{2}{3} \Theta^2 = 2 \kappa \rho
\end{equation}
where we defined $ ^\ast\!R : =~ ^3_1R$.

\item{$\bullet$} Momentum constraint equation
\begin{equation}
\partial_i \Theta = 0\ .
\end{equation}
\end{description}

\medskip
\noindent
Constraint equations (10), (11)
\begin{description}
\item{$\bullet$} 
\begin{equation} \partial_i \rho = 0;
\end{equation}
\item{$\bullet$} 
\begin{equation} 
\partial_i q_j - \frac{1}{3} \delta_{ij} \partial_e q_e = 0; 
\end{equation}
\item{$\bullet$} Local energy conservation equation
\begin{equation}
D_t \rho + \Theta\rho + \partial_e q_e = 0;
\end{equation}
\item{$\bullet$} Propagation equation for the energy flow vector
\begin{equation}
D_t q_i + \frac{4}{3} \Theta q_i = 0.
\end{equation}
\end{description}

Furthermore we consider eq. (2) which reads, in the nonrelativistic shear-free limit in Galilean coordinates,
\begin{equation}
\partial_i u_j - \frac{1}{3} \delta_{ij} \Theta = 0 \ .
\end{equation}
It is now an easy task to determine the dependence of $\Theta, \rho, u_i$ and $q_i$ on the space coordinates $\vec{x}$. From (29) and (30) respectively we infer that the volume expansion $\Theta$ and the energy density $\rho$ are functions of time $t$ only. Therefore, due to (28) and (32), respectively, the same holds for $^\ast\!R$ and $\partial_e q_e$. Then, with the definition $q(t) : = \frac{1}{3} \partial_e q_e$, we obtain from (31)    
\begin{equation}
q_i = x_i q(t)
\end{equation}                                                                                                                      
and (34) is solved by 
\begin{equation}
u_i = x_i \frac{\Theta (t)}{3} \ .
\end{equation}                                                                                                                                                               
Insertion of (35) and (36) into (33) leads to a differential equation for q(t),
\begin{equation}
\dot{q} + \frac{5}{3} \Theta q = 0 \ .
\end{equation}                                                                                                                                 
Next we define the cosmic scale factor $a(t)$ by 
\begin{equation}
\frac{\dot{a}}{a} : = \frac{1}{3} \Theta (t) \ .
\end{equation}

Then we get the well-known form  for the RE- equation (27),
\begin{equation}
\ddot{a} = - \frac{\kappa a \rho}{6}
\end{equation}
and the local energy conservation equation (32) reads
\begin{equation}
\dot{\rho} + 3 \frac{\dot{a}}{a} \rho + 3 q = 0 \ .
\end{equation}
Now we are able to relate the curvature $^\ast\!R(t)$  to the function $q(t)$ .
For that we define a curvature function $K(t)$ by
\begin{equation}
K(t) : = \frac{a^2}{6}~ {^\ast\!R},
\end{equation}
whose time derivative is proportional to $q(t)$, 
\begin{equation}
\dot{K} = - \kappa q a^2 
\end{equation}
as it can easily be seen with the aid of(28) and (38) - (40).

So, in the absence of an energy flow vector, we would have $\dot{K}  = 0$ and therefore $K =$ const. (FLRW Universe).

\section{Large-scale cosmological model}

Summarizing the foregoing results, we define the nonrelativistic cosmological model at large scales by the following system of three coupled ordinary differential equations (ODEs) for the cosmological scale factor, $a(t)$, the active gravitational mass density,   $\rho (t)$,  (we define $\hat{\rho} : = \frac{\kappa a^3 \rho}{6}$), and the energy flow vector $q_i = x_i q(t)$
\begin{equation}
\ddot{a} = - \frac{\hat{\rho}}{a^2} \ ,
\end{equation}             
and
\begin{equation}
\dot{\hat{\rho}} + \frac{\kappa}{2} q a^3 = 0
\end{equation} 
and
\begin{equation}
\dot{q} + 5 \frac{\dot{a}}{a} q = 0
\end{equation}
with  curvature function $K(t)$ given by
\begin{equation}
K = - \dot{a}^2 + \frac{2\hat{\rho}}{a} \ .
\end{equation}

Equations (43), (44) and (45) are equivalent to the cosmological equations derived for the first time in [10], section V, obtained by restricting the solutions of the EOMs for the darkon fluid model to such which obey the cosmological principle [10].  

Due to their nonrelativistic nature, these equations contain no curvature function; eq. (46) resides outside the darkon fluid model. But, as we have shown in Sect. 4, a nontrivial curvature function can be incorporated into a nonrelativistic framework by  taking into account the 1st-order relativistic correction of the spatial curvature. This led us to Eq. (46).

We remark that such a connection between mass density and curvature, as given by (46), has  already been postulated in a static nonrelativistic context in a recent paper by Abramowicz et al. [22] (called ``Enhanced Newtonian Gravitational Theory'').

\bigskip
\noindent
{\bf 5.1 Dynamics and symmetries in minisuperspace}

\medskip
\noindent
To gain deeper understanding of the Equations (43) - (46) and to find their possible symmetries we should derive them from a variational principle. We succeeded  doing that only after having eliminated the function q(t) by integrating (45), 
\begin{equation}
q(t) = \frac{2 K_1}{\kappa a^5 (t)},
\end{equation}
where the constant $K_1$, which has been chosen in accordance with [13], is fixed by the initial value of the energy flow vector $q_i$. By substituting (47) for $q$ in (44) yields the cosmological equations in reduced configuration space
\begin{equation}
\ddot{a} = - \frac{\hat{\rho}}{a^2}
\end{equation}             
\begin{equation}
\dot{\hat{\rho}} + \frac{K_1}{a^2} = 0
\end{equation}
and
\begin{equation}
K = - \dot{a}^2 + \frac{2\hat{\rho}}{a} \ .
\end{equation}
For the time derivative of the curvature function  we obtain from (50) due to (48) and (49) (cp. (42))
\begin{equation}
\dot{K} = - \frac{2K_1}{a^3}\ .
\end{equation}
So the constant $K_1$ is a measure of the strength of the time variation of the curvature function.
            
Let us now consider the following $1^{st}$-order minisuperspace Lagrangian $L$
\begin{equation}
L = \frac{K \dot{\hat{\rho}}}{2 K_1} - \frac{\dot{a}b}{a^2} - H
\end{equation}
with the Hamiltonian $H$ given by
\begin{equation}
H = N H_c , ~~~~~~~~~ H_c = - \frac{1}{2}\left( \frac{b}{a}\right)^2 - \frac{K}{2a^2} + \frac{\hat{\rho}}{a^3}\ .
\end{equation}
Here $N$ is the lapse function which takes care of the time-reparametrization invariance of the action $S = \int dt L$.

Variation of the Lagrangian (52) w.r.t. a, b, K, $\hat{\rho}$ and $N$ leads to the four EOMs
\begin{eqnarray}
\dot{b} &=& - \frac{N \hat{\rho}}{a^2} - 2 a H\\
\dot{a} &=& Nb\\
\dot{\hat{\rho}} &=& - \frac{N K_1}{a^2}\\
\dot{K} &=& - \frac{2 N K_1}{a^3}
\end{eqnarray}
and the Hamiltonian constraint
\begin{equation}
H = 0.
\end{equation}                                                                                                                                                                            
In the gauge $N = 1$ (choice of proper time) we obtain on the constrained surface $H = 0$ for (54) - (57) just the original cosmological Equations (48)-(51).

To recast the Euler-Lagrange Equations (54) - (57) into the Hamiltonian form, we use,  instead of the canonical one, the symplectic approach.

Then, in the gauge $N = 1$ chosen here and in the following, eqns (54) - (57) take the form, where $q_i \in (a,b,K,\hat{\rho})$ are the coordinates in a 4-dimensional generalized phase-space,
\begin{equation}
\dot{q}_i = \sum_j [q_i,q_j] \frac{\partial H}{\partial q_j},
\end{equation}
with  nontrivial Poisson-brackets (PBs) $[q_i,q_j]$,                                  
\begin{equation}
[a,b] = - a^2 \qquad \mbox{and} \qquad [K, \hat{\rho}] = - 2 K_1\ .
\end{equation}

Now we  inquire about  dynamical (hidden) symmetries of the system of our EOMs (59). Let us define two functions $Q_{2,3}$ on generalized phase-space [13]
\begin{equation}
Q_2 : = K_1 b - \frac{1}{2} \hat{\rho}^2
\qquad\hbox{and}\qquad
Q_3 : = - \frac{\hat{\rho}^3}{6} - Q_2 \hat{\rho} + \frac{K^2_1}{a}\ .
\end{equation}
From the EOMs (59) we obtain for their time-derivatives 
\begin{equation}
\dot{Q}_2 = - 2 K_1 a H
\qquad\hbox{and}\qquad
\dot{Q}_3 = 2 K_1 a \hat{\rho} H\ .
\end{equation}
Their mutual PB yields a constant,
\begin{equation}
[Q_2, Q_3] = - K^3_1\ .
\end{equation}
We observe that $Q_{2,3}$ are conserved only on the constrained surface $H = 0$, hence they represent conditional symmetries [23].

Let us now consider the infinitesimal transformations of the $q_i$ and the corresponding velocities $\dot{q}_i$ generated by $Q_k ~~(k = 2, 3)$
\begin{equation}
\delta_k q_i = \epsilon [q_i, Q_k]
\end{equation}
and                
\begin{equation}
\delta_k \dot{q}_i = \epsilon [\dot{q}_i, Q_k] = \epsilon \left[ [q_i H], Q_k\right]\ .
\end{equation}
Calculating (65) requires care: Because of $\dot{Q}_k \ne 0$ in the unconstrained generalized phase space, we will have
$$
\delta_k \dot{q}_i \ne (\delta_k q_i)^\cdot\ .
$$
Explicitly, using the Jacobi identity for the r.h.s. of (65) we obtain
\begin{equation}
\delta_k \dot{q}_i = - \epsilon [q_i, \dot{Q}_k] + (\delta q_i)^\cdot\ .
\end{equation}
It is straightforward  to compute the explicit expressions for the infinitesimal transformations of the $q_i$ (64) and the $\dot{q}_i$ (66). We will dispense with that.  
But we note that, surprisingly, they leave the EOMs (54) - (57) as well as the PBs (60) invariant without producing any additional terms proportional to $H$. 

Additionally we note that the time-reparametrization invariance of the action leaves, in the gauge $N = 1$ and on the constrained surface $H = 0$, the dilational symmetry as residual symmetry. 
Indeed the cosmological Equations (48) - (51) are invariant w.r.t. the transformations $(\lambda \in {\bf R}^1)$
\begin{equation}
t \to t^\ast = \lambda t, ~~~~~~~~~ q_i (t) \to q_i^\ast (t) = \lambda^{z_i} q_i (\lambda t)
\end{equation}
where the scale dimensions $z_i$ of the generalized phase space coordinates $q_i$ are given by
$$
z_a = - 3/5, ~~~~ z_b = 2/5, ~~~~ z_{\hat{\rho}} = 1/5~~~and~~~z_K = 4/5\ .
$$                                                                                                                                                                                       
Then, in the scaling limit,
\begin{equation}
q_i = q^\ast_i \qquad \mbox{we~obtain} \qquad q_i (t) \sim t^{-z_i}.
\end{equation}

\bigskip
\noindent
{\bf 5.2 Solution of the cosmological Equations}

\medskip
The (physical) solution space is defined as the common space of solutions of the constraint (50) and of the EOMs (48), (49). As shown by (62), the functions $Q_k$ (61) take constant values $K_k$ on this solution space.  So (48) and (49) are completely integrable: 

\medskip
\noindent 
First of all, the expression for $Q_3$ (61) gives the algebraic Equation  $(a_t : = K^2_1/K_3$ is the transition scale factor)
\begin{equation}
\frac{\hat{\rho}^3 (a)}{6} + K_2 \hat{\rho} (a) + K_3 \left( 1 - \frac{a_t}{a}\right) = 0,
\end{equation}
whose solution leads to an explicit expression for $\hat{\rho}$ as a function of the scale factor $a$ [10]
$$
\hat{\rho} (a) = u_+ (a) + u_- (a)
$$ 
with
\begin{equation}
u_\pm (a) : = \left( - 3 v (a) \pm [(2 K_2)^3 + (3 v (a))^2]^{1/2} \right)^{1/3}
\end{equation}                                                                                                                                                                          
and
$$
v (a) : = K_3 (1-\frac{a_t}{a})\ .
$$
Next, the expression for $Q_2$ (61) leads to a separable $1^{st}$-order ordinary differential equation (ODE), solved as,
\begin{equation}
t - t_0 = K_1 \int \frac{da}{K_2 + \frac{1}{2} \hat{\rho}^2 (a)}\ .
\end{equation}
The integral on the r.h.s. of (71) can easily be calculated without using the explicit expression (70) for $\hat{\rho}(a)$. By means of (69) we may rewrite it as [10] ;       
\begin{equation}
t - t_0 = - K_1 K_3 a_t \int \frac{d \hat{\rho}}{( \frac{1}{6} \hat{\rho}^3 + K_2 \hat{\rho} + K_3)^2}
\end{equation}
leading to a lengthy expression in terms of elementary functions. We omit the details and refer to [10], appendix A.

A very important quantity for cosmological considerations is the Hubble function $H : = \dot{a}/a$, which is easily obtained from $Q_2$ (61) in terms of $\hat{\rho} (a)$ and $a$
\begin{equation}
H(a) = \frac{1}{K_1 a} \left( K_2 + \frac{1}{2} \hat{\rho}^2 (a)\right) \ .
\end{equation}
We summarize: Equations (70) and (72) determine the active gravitational mass density $\rho(t)$ and the scale factor $a(t)$ in terms of three integration constants $K_1, K_2$ 
and $K_3$ (we recall that $K_1$ is determined by the initial value of the energy flow vector).

To get a deceleration/acceleration transition at $a = a_t > 0$, we conclude from (48) and (69) that [10], [13]
\begin{itemize}
\item $K_{2,3}$  take necessarily positive values $K_{2,3} > 0$, leading,  due to the first formula in (61), to $K_1 > 0$ for an expanding Universe. 
\item Then for $a < a_t$ we are in the decelerating phase and for $a > a_t$ we are in the accelerating phase of the Universe.
\end{itemize}

In our former work [10], [13] we have also compared the predictions of our model for the Hubble function as a function of the redshift $z := \frac{1}{a} - 1$ (see the next section) with observational data of $H(z)$. Thereby we determined the three integration constants $K_i$ by some observational values for $z_t$, the Hubble parameter $H_0 : = H(0)$ and the deceleration parameter.
                              
Now we inquire about what is  new when compared to our former work [10], [13].
In fact, by considering our nonrelativistic model as a limit of a relativistic one, we found a new physical interpretation
:
\begin{description}
\item{--} The model contains no negative pressure. Hence any kind of dark energy is absent.
\item{--} The model possesses a time-dependent spatial curvature $^\ast\!R(t)$ which is dynamically determined by known values of the active gravitational mass density $\rho$ and the cosmic scale factor $a$
\item{--} The deceleration$/$acceleration transition, given by a sign change of $\rho$, and the subsequent acceleration phase of the Universe are driven by the time derivative of the curvature function.
\item{--} The conserved charges $Q_{2,3}$ represent conditional dynamical symmetries of the cosmological Equations. 
\end{description}

Similar results have been  obtained  recently in the relativistic darkon fluid framework [9] at sub- Hubble scales. Now we have learned that these findings are valid generally, as far as a nonrelativistic description of the dark sector of the Universe is permitted.

To express the curvature function  in terms of the scale factor, $a$, we insert the expression for $H(a)$ (73) into (46) and obtain,
\begin{equation}
K(a) = \frac{2 \hat{\rho} (a)}{a} - \frac{1}{K_1^2} \left( K_2 + \frac{1}{2} \hat{\rho}^2 (a)\right)^2 \ .
\end{equation}      

We conclude~:
\begin{description}
\item{--} For the present \emph{accelerating phase} of the Universe $K(a)$ turns out to be negative (due to (43) we have $\hat{\rho} < 0$). So we get a \emph{hyperbolic space}. 
\item{--} For the early \emph{decelerating phase} $( a\ll 1)$ we obtain from (69)
\begin{equation}
\hat{\rho} (a) \simeq \left( \frac{6 K^2_1}{a} \right)^{1/3}
\end{equation}
leading, due to (74), to
\begin{equation}
K(a) \simeq \frac{1}{2a} \left( \frac{6 K^2_1}{a} \right)^{1/3}\ .
\end{equation}
So we get a \emph{spherical space}.

Note that the  relations (75) and (76) are exact in the scaling limit (68) we obtain for $K_2 = K_3 = 0$.                   

\item{--} The transition from a spherical to a hyperbolic space will take place at some $a = a_0$ with $a_0 < a_t$.
\end{description}

\section{Light propagation}

The comparison of our model with observational data in [10] and  [13] rests upon the use of model-independent data for the Hubble function $H(z)$ derived from differential ages of galaxies (cosmic chronometer method, see [24]). On the other hand; the supernovae SN Ia-data (see the Union 2.1 compilation [25]) are given in terms of the distance modulus defined by the logarithm of the luminosity distance  $D_L$         
which, by the reciprocity relation (cp. [15]), is universally related to the angular diameter distance $D_A$ ( called ``area distance'' in [15]) by,
\begin{equation}
D_L = (1 + z)^2 D_A \ .
\end{equation}
The redshift $z$, which will be the independent variable here and in the following, is simply related to the cosmic scale factor $a$ if the shear vanishes,
\begin{equation}
1 + z = a^{-1}\ 
\end{equation}
(cp. [26]).
Now the point to be made is that the luminosity distance $D_L$ or, by (77), the area distance $D_A$ depends heavily on the underlying cosmological model. The redshift dependence of the curvature function $K(z)$,  generated by the nontrivial energy flow vector in our model, does not allow to use the FLRW-formula for $D_A$.

For a general cosmological model the area distance $D_A$  satisfies a differential equation ([27], eq. (49) with (32) and (46)) which, if applied to our case, is given by 
\begin{equation}
(1+z)^2 H^2 D_A^{\prime\prime} + (1+z) (2H^2 - \dot{H}) D_A^\prime = - \frac{\kappa}{2} (\rho + 2 q^\mu e_\mu ) D_A
\end{equation} 
($D_A^\prime : = \frac{d}{dz} D_A$),
where $e^\mu$ is the spatial direction of observation satisfying $e^\mu e_\mu = 1$ and in the NRL $n_\mu e^\mu = 0$. 
Then in our case $e^\mu = (0,1)$ is the radial direction in spherical co--moving coordinates. 
In the same frame the energy flow vector takes according to (35) and (47) the form
\begin{equation}
q^\mu = \left( 0, \frac{2 K_1}{\kappa a^4} r\right),
\end{equation}
where we have used the relation $\vec{x} = a \vec{r}$  between Galilean coordinates  $\vec{x}$ given in the Eulerian frame and co-moving coordinates $\vec{r}$.

So we obtain 
\begin{equation}
(1+z)^2 H^2 D_A^{\prime\prime} + (1+z) \left( 2H^2 + (1+z) H H^\prime \right) D_A^\prime = 
- \left( \frac{\kappa}{2} \rho + 2r(z) K_1 (1+z)^4 \right) D_A,
\end{equation}                                                                                                                                                    
where $r(z)$ is the radial distance between observer and the galaxy at redshift $z$
\begin{equation}
r(z) : = \int^z_0 \frac{dz^\prime}{H(z^\prime)}
\end{equation}
and we have used for a generic function  $A(z)$          the relation
\begin{equation}
\dot{A} (z) = - (1+z) H A^\prime (z) \ .
\end{equation}
With the Ansatz $D_A = (1 + z)^{-1} f (z)$                                                                                                                                                                 
we obtain for $f(z)$ the differential equation,
\begin{equation}
Hf^{\prime\prime} + H^\prime f^\prime = - \left( \frac{3\hat{\rho} (1+z)}{H} - \frac{H^\prime}{1+z} + \frac{2 K_1 r(z) (1+z)^2}{H} \right) f\ .
\end{equation}
 
The r.h.s of (84) turns out to be 
\begin{equation}
\mbox{r.h.s.~(84)}~ = - (r(z) K(z))^\prime f \ .
\end{equation}                                                                                                                 

To prove (85) we observe that (43) and (46) may be rewritten as
\begin{equation}
\frac{\hat{\rho} (1+z)}{H} = \frac{H^\prime}{1+z} - \frac{H}{(1+z)^2} ~~~\mbox{and}~~~ K(z) = - \frac{H^2}{(1+z)^2} + 2 \hat{\rho} (1+z) \ .
\end{equation}
By combining the relations (86) with (51) and (84) we get immediately (85).
So we have finally,
\begin{equation}
(Hf^\prime)^\prime + (r K)^\prime f = 0,
\end{equation}
which has to be solved with initial conditions $f(0) = 0, f^\prime (0) = 1/H(0)$.

We did not succeed to obtain an analytic solution for (87). Therefore (87) has to be treated by numerical methods.

For $K =$ const. the solution of (87) is given by the well-known FLRW-formula
$$
f(z) = \frac{1}{\sqrt{K}} \sin (\sqrt{K} r (z))\ .
$$

For the sake of completeness we notify a cubic equation for the Hubble function $H(z)$ which is easily obtained by means of (69) and (73),
\begin{equation}
(K_3 - K^2_1 (1+z))^2 = \frac{2}{9} \left( \frac{K_1 H(z)}{1+z} - K_2 \right) \left( \frac{K_1 H(z)}{1+z} + 2 K_2 \right)^2\ .
\end{equation}
 
\section{Conclusions}

The question arises : how does a model which predicts a non-vanishing and  redshift-dependent spatial curvature fit with observational data? It has been argued in most of the recent papers that the data strongly support spatial flatness (cp. [2]). So let us give some counter-arguments:
\begin{description}
\item{$\bullet$} There is no model-independent evidence for spatial flatness. Almost all papers supporting the idea are based on the standard cosmological $\Lambda$CDM-model.
But the very possibility of a time-dependent evolution of the spatial curvature is absent within this model [5].

\item{$\bullet$} Stephani models, which contain a time-dependent spatial curvature, are also able to explain the accelerated expansion of the Universe [28], [29].

\item{$\bullet$} Theoretical arguments which say that a small departure of the metric from  standard background implies  small spatial curvature are misleading, due to possible strong second derivatives (J. Ehlers (2007), see the memorial article by T. Buchert et al. [30]).

\item{$\bullet$} For cosmological models based on averaging over the inhomogeneities of the Universe, one comes to the conclusion that the present day cosmic acceleration is due to  negative average spatial curvature [31], [32]. In particular Roukema et al. conclude, in a very recent paper [33]:

``Pending more accurate, relativistic calculations, it would seem prudent to consider ``dark 
energy'' as an artifact of virialisation-induced negative spatial curvature and void-dominated expansion rates $\ldots$''
\end{description}

In this context  Larena et al. [34] suggested  to introduce an effective metric which mimics the geometry of the averaged Universe. For this so called ``template metric'' the authors of [34] propose an Ansatz which differs from the standard FLRW-metric only by the time-dependence of the curvature function K(t). But for such a metric the EFEs show necessarily a nonzero $tr$-component of the Einstein tensor [35] and, therefore, it is mandatory for the EMT to contain a nonvanishing energy flow vector. So it seems not to be absurd to consider our cosmological model as a first step in the search for a cosmological model with such a template metric.
                                                   
                                                                                                                                                            Another point to be made is the violation of at least the weak energy condition in our model (we have  $\rho > 0$  in the decelerating phase but $\rho < 0$ in the accelerating phase of the Universe). But the long-standing believe that energy constraints on the EMT are necessary for a relativistic fluid to be physically reasonable are on the verge of dying [36] (e.g. quantum corrections for the coupling of scalar fields to gravity violate all local energy conditions [36]). There is no reason to worry about this point therefore. 

The solutions of our cosmological equations depend on three positive constants $K_i (i = 1, 2, 3)$ which are related to the initial conditions for $\Theta$, $\rho$ and $\dot{K}$ (or $\dot{\rho}$). To decide whether our model is an admissible alternative to the standard cosmological model or not, a least-squares fit to the cosmic chronometer data [24] (a preliminary comparison was  made in [10] and [13]) and for the SN Ia-data (Union 2.1-compilation [25])  is called for. 
 The latter case requires a numerical treatment of the differential equation (87) for the area distance (but such a numerical work is beyond the means of the present author).

In this paper we have treated exclusively the behavior of the Universe at large scales. It is an easy task to derive within our cosmological model the nonrelativistic dynamical equations also for the case of a nonvanishing Weyl tensor which, however, will contain nontrivial shear. This gives a basis for considering structure formation and other phenomena on astrophysical scales. To some extent this has already been done in our former work. In [9] and [13] we have shown that our model describes successfully the flat behavior of galactic rotation curves. 

\section*{Acknowledgements} I'm grateful to Thomas Buchert for valuable hints and correspondence. I thank the referee for his constructive criticism and Peter Horvathy for very helpful remarks.

\section*{References}

\begin{description}
\item{[1]} M. Moresco et al., arXiv: 1601.01701
\item{[2]} J. Adamek et al., arXiv: 1512.05356
\item{[3]} Th. Burchert et al., arXiv: 1512.03313
\item{[4]} A. Padilla, arXiv: 1502.05296
\item{[5]} Th. Buchert and S. R\"as\"anen, Ann. Rev. Nucl. Particle Sci.62, 57 (2012)
\item{[6]} K. Koyama, arXiv: 1504.04623
\item{[7]} J. Moffat, arXiv: 1510.07037
\item{[8]} S. Tsujikawa, Class.  Quant. Grav. 30, 214003 (2013). arXiv: 1304.1961
\item{[9]} P. Stichel and W. Zakrzewski, Eur. Phys. J. C (2015) 75:9. arXiv: 1409.1336v2
\item{[10]} P. Stichel and W. Zakrzewski, Phys. Rev. D 80, 083513 (2009). arXiv: 0904.1375
\item{[11]} P. Stichel and W. Zakrzewski, Eur. Phys. J. C (2010) 70:713. arXiv: 1008.1200
\item{[12]} P. Stichel and W. Zakrzewski, Int. J. Geom. Meth. Mod. Phys. 9, 1261014  (2012).\\
          arXiv: 1202.4895
\item{[13]} P. Stichel and W. Zakrzewski, Entropy 15, 559 (2013). arXiv: 1301.4486 
\item{[14]} G. Ellis and H. Van Elst, arXiv:  gr-qc/9812046v5
\item{[15]} G. Ellis et al., Relativistic cosmology, Cambridge University Press, Cambridge 2013
\item{[16]} J. Senovilla et al., Gen. Rel. Grav. 30, 389 (1998). arXiv:  gr-qc/9702035
\item{[17]} A. Coley et al., Class. Quant. Grav. 11, 1261 (1994). arXiv: gr-qc/9405034
\item{[18]} L. Herrera et al., Gen. Rel. Grav. 42, 1585 (2010).arXiv: 1001.3020
\item{[19]} E. Cartan, Ann. Ec. Norm. Sup. 40, 325 (1923) and 41, 1 (1924)
\item{[20]} G. Dautcourt, Acta Phys. Pol. XXV, 637 (1964) and B 21,  755 (1990)
\item{[21]} H. K\"unzle, Gen. Rel. Grav. 7, 445 (1976)
\item{[22]} M. Abramowicz et al., Gen. Rel. Grav. 46, 1630 (2014)
\item{[23]} K. Kuchar, J. Math. Phys. 23, 1647 (1982);\\
         T. Christodoulakis et al., J. Geom. Phys. 71, 127 (2013)
\item{[24]} M. Moresco, arXiv: 1503.01116
\item{[25]} N. Suzuki et al., Astrophys. J. 746, 85 (2012). arXiv: 1105.3470
\item{[26]} S. R\"as\"anen, JCAP 0902:11 (2009). arXiv: 0812.2827
\item{[27]} C. Clarkson et al., MNRAS 426, 1121 (2012). arXiv: 1109.2484
\item{[28]} W. Godlowski et al., Class. Quant. Grav. 21, 3953 (2004)
\item{[29]} A. Balcerzak et al., Phys. Rev D 91, 083506 (2015)
\item{[30]} T. Buchert et al., Gen. Rel. Grav. 41, 2017 (2009). arXiv: 0906.0134
\item{[31]} T. Buchert, Gen. Rel. Grav. 40, 467 (2008)
\item{[32]} T. Buchert and M. Carfora, Class. Quant. Grav. 25, 195001 (2008)
\item{[33]} B. Roukema et al., JCAP, 10, 043 (2013). arXiv: 1303.4444
\item{[34]} J. Larena et al., Phys. Rev. D 79, 083011 (2009)
\item{[35]} S. R\"as\"anen, Astropart. Phys. 30, 21 (2008)
\item{[36]} C. Barcelo and M. Visser, Int. J. Mod. Phys. D 11, 1553 (2002); arXiv: gr-qc/0205066

\end{description}

\end{document}